\documentclass[
]{ceurart}

\usepackage{listings}
\usepackage{graphicx}
\usepackage{subcaption}
\begin{document}

\copyrightyear{2022}
\copyrightclause{Copyright for this paper by its authors.
  Use permitted under Creative Commons License Attribution 4.0
  International (CC BY 4.0).}

\conference{GREEN-AI: Grounding Research on Energy- and resource-Efficient Networks for sustainable AI,
  November 08, 2026, Rome, IT}

\title{EMAR: Efficient Multi-domain Adaptive Retrieval}


\author[1]{Marco Braga}[%
email=marco.braga1@unimib.it,
]
\cormark[1]
\address[1]{University of Milano-Bicocca, Italy}
\author[1]{Pranav Kasela}[%
email=pranav.kasela@unimib.it,
]

\author[2]{Ophir Frieder}[%
email=ophir@ir.cs.georgetown.edu,
]
\address[2]{Georgetown University, USA}

\author[2]{Nazli Goharian}[%
email=nazli@ir.cs.georgetown.edu,
]

\author[1]{Gabriella Pasi}[%
email=gabriella.pasi@unimib.it,
]

\author[3]{Raffaele Perego}[%
email=raffaele.perego@isti.cnr.it,
]
\address[3]{ISTI-CNR, Italy}



\cortext[1]{Corresponding author.}

\begin{abstract}
Neural retrieval systems are increasingly deployed in multi-domain settings, where each domain may have its own vocabulary and word distribution. A common approach is to fine-tune separate models for each domain, achieving strong in-domain performance while increasing training costs and storage requirements. In contrast, a single model trained jointly across domains avoids model duplication but weakens robustness under domain shift. This creates a trade-off between effectiveness and resource-efficiency. To address this issue, we propose EMAR, a retrieval system that employs a single frozen model shared across all domains and introduces lightweight domain-specific adapters. At inference time, a learned routing mechanism dynamically selects the most suitable adapter for each query. EMAR enables efficient domain-specific retrieval by sharing the computationally expensive backbone across multiple domains and without maintaining multiple full models. 
Experimental results on three multi-domain benchmarks spanning seventeen domains and three retrieval architectures show that EMAR achieves retrieval effectiveness on par or even better than fully fine-tuned domain-specific models while requiring storing only a small fraction of parameters per domain. Furthermore, EMAR exhibits strong generalization to previously unseen domains, highlighting its effectiveness as an extensible solution for multi-domain retrieval.

\end{abstract}


\maketitle

\label{sec:introduction}

Recent advances in pretrained 
and Large Language Models (LLMs) significantly improve the effectiveness of neural information retrieval systems~\cite{zhu2023large}. In particular, transformer-based encoders enable retrieval models to learn rich semantic representations of queries and documents, outperforming traditional sparse lexical retrieval methods across a wide range of tasks~\cite{reimers19sentence,devlin-etal-2019-bert, karpukhin-etal-2020-dense}.  
Within this paradigm, both bi-encoder~\cite{Survey_dense__acm_tois} and cross-encoder~\cite{nogueira20passage} retrieval architectures become dominant approaches for semantic search due to their strong representational capacity~\cite{zhu2023large}.
In practice, modern neural retrieval systems are increasingly required to work across multiple heterogeneous domains, including scientific literature, web documents, and social media platforms~\cite{zhang2025pmta}. A common strategy is to train retrieval models on a large-scale dataset, such as MS MARCO~\cite{MSMARCO}, and apply them to downstream domains in a zero-shot setting. However, prior works~\cite {thakur2021beir, vast2024simple}
show that retrieval effectiveness can degrade under domain shift due to differences in vocabulary, writing style, and semantic distributions. As a result, domain-specific fine-tuning is often needed to achieve strong in-domain retrieval performance.
A popular solution is to train a separate retriever for each domain~\cite{Goli_medical_dense_sigir, gueta23knowledge}. Parameter-efficient tuning methods such as Houlsby~\cite{houlsby19parameter} and LoRA~\cite{hu2022lora} adapters can reduce the cost of such domain specialization without requiring full model retraining.
While domain-specific retrievers generally provide strong effectiveness, this approach scales poorly as the number of domains increases, since each domain requires training, storing, and maintaining an independent set of model parameters. 
Another complementary direction investigates knowledge distillation to transfer knowledge from strong teacher models to compact retrievers~\cite{hinton15distilling, hofstätter21improving, lin21batch, qu21rocketqa}. 
An alternative approach involves training a single unified retriever jointly across multiple domains~\cite{mysore2025bridging}. Such models improve deployment efficiency through parameter sharing, but they underperform domain-specialized retrievers due to interference between heterogeneous data distributions and domain mismatch~\cite{thakur2021beir}. 

To address this trade-off, we propose EMAR (Efficient Multi-domain Adaptive Retrieval), a modular framework for efficient multi-domain neural retrieval.
EMAR combines a single frozen backbone shared across all domains with lightweight domain-specific adapters that can be trained and deployed independently. For each domain, we train a lightweight adapter through knowledge distillation~\cite{morand2026reproducing} to reproduce the relevance estimation behaviour of a domain-specific teacher retriever while simultaneously learning from supervised relevance judgments. 
While prior works~\cite{hinton15distilling, hofstätter21improving, lin21batch, qu21rocketqa} primarily uses distillation for general model compression, we leverage it to inject domain-specific behaviour into lightweight adapters while keeping the shared backbone frozen. 
This design enables EMAR to efficiently incorporate new domains without retraining the full retrieval model.
In addition, EMAR introduces a learned routing mechanism that dynamically selects the most suitable domain adapter during inference. The selected adapter, together with the frozen shared backbone, is then used to compute relevance scores for ranking. This conditional computation strategy preserves domain specialization while avoiding the computational cost of executing multiple full retrievers during inference. 
The gating function and the adapter selection take inspiration from the Mixture-of-Expert (MoE) architecture with a top-1 gate routing mechanism \cite{jacobs91adaptive}. In MoE, the training of the whole architecture is performed end-to-end. Our approach diverges from MoE~\cite{feng2024mixture,yang2024mlora,song2024multilora} by training the various components independently from each other, enhancing the scalability of the training procedure.
Adding a new domain requires only training a new domain-specific adapter module and the classifier, while all previously learned components remain frozen. However, we show that the learned router generalizes effectively to previously unseen domains, enabling EMAR to incorporate new domain adapters without the need of updating the routing mechanism and thereby without introducing additional learned components.
We evaluate EMAR on three multi-domain retrieval benchmarks spanning seventeen domains and three retrieval architectures. Experimental results show that EMAR consistently achieves performance on par with, and sometimes better than, fully fine-tuned domain-specific retrievers while requiring training and storing only a small fraction of parameters per domain. Furthermore, EMAR significantly outperforms unified multi-domain baselines, reduces computational overhead compared to multiple domain-specific models, and shows strong generalization to previously unseen domains without additional retraining of the classifier. These findings highlight the effectiveness of EMAR as an efficient system for multi-domain neural retrieval.
Our contributions can be summarised as follows: (i) We propose EMAR, a modular system for multi-domain neural retrieval that separates domain specialization from the retrieval backbone via lightweight adapters; (ii) We show that EMAR matches or even improves upon strong domain-specific retrievers across multiple multi-domain benchmarks while significantly reducing computational overhead.

\begin{figure}[t]
   \centering
    \includegraphics[width=\linewidth]{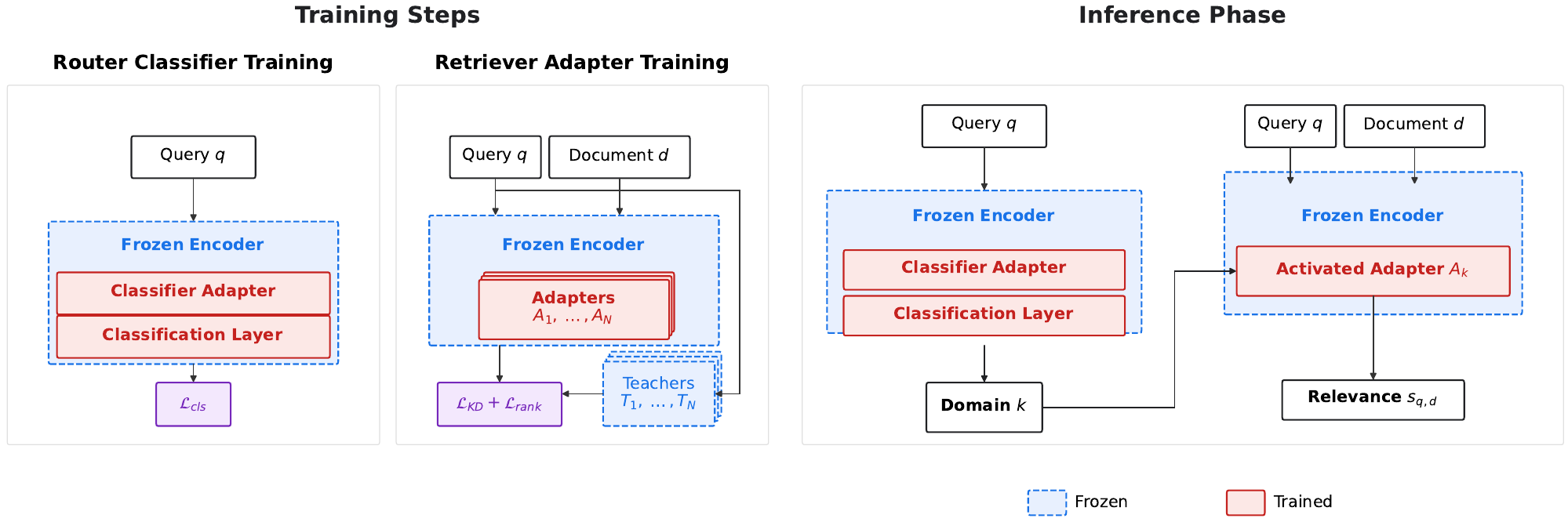}
    \caption{Overview of EMAR training (on the left) and inference (on the right).}
    \label{fig:distillationmodel}
\end{figure}

\section{EMAR}

In this section, we introduce EMAR (Efficient Multi-domain Adaptive Retrieval), whose overall framework is shown in Figure~\ref{fig:distillationmodel}. EMAR is designed to address the limitations of retrieval systems operating across multiple domains, where deploying separate domain-specific models incurs substantial computational overhead.
Let $\mathcal{D}=\{d_1,\dots,d_N\}$ denote a set of different domains. In conventional multi-domain settings, each domain is typically handled either by a multiple domain-specific retrievers or by a single unified model shared across all domains. The former scales poorly with the number of domains, while the latter often suffers from performance degradation due to conflicts across heterogeneous data sources.
To address this trade-off, EMAR relies on a set of lightweight domain-specific adapters $\{A_i\}_{i=1}^N$ attached to a frozen shared retriever $f_\theta$. 
Each adapter $A_i$ is trained independently through retrieval-specific supervision and knowledge distillation from a domain-specialized teacher $T_i$.
At inference time, a learned router mechanism activates one adapter per query, yielding a single forward pass whose cost does not grow with the number of supported domains. 
This design makes EMAR particularly suitable for incremental multi-domain retrieval, where new domains can be incorporated by adding compact adapter modules without modifying the shared backbone or existing adapters.


\subsection{Domain-Specific Adapters}

EMAR assigns to each domain $d_i$ a lightweight adapter $A_i$, inserted into the shared encoder to model domain-specific knowledge. The adapter is responsible for learning domain-dependent ranking signals that are not fully represented in the general-purpose encoder.
Importantly, for each domain, only the parameters of $A_i$ are updated, while the shared encoder $f_\theta$ remains fixed, ensuring that each domain can be added independently without affecting existing ones. 
We assume access to a domain-specific teacher retriever $T_i$, obtained via full fine-tuning on labeled data. EMAR leverages knowledge distillation to transfer the ranking behaviour of $T_i$ to $A_i$, while also using a ranking loss over the relevance judgments. The distillation loss $\mathcal{L}_{KD}$ aligns student and teacher relevance scores and is defined as the MSE~\cite{kim2021comparing} between the teacher and student similarity scores. The ranking loss $\mathcal{L}_{rank}$ is implemented as a Triplet Margin loss~\cite{balntas2016learning} and forces the model to distinguish relevant from non-relevant documents. Following common practice in dense retrieval~\cite{morand2026reproducing}, the final loss is defined as
$\mathcal{L} = \mathcal{L}_{KD} + \mathcal{L}_{rank}$.

\subsection{Routing Mechanism}

At inference time, EMAR must dynamically select the appropriate adapter for each query without requiring explicit domain labels. To this end, we introduce a routing function $G(\cdot)$ that predicts the most likely domain given an input query $q$:
$G(q) = \arg\max_i P(d_i \mid q)$.
The router is implemented as a lightweight adapter and a classifier on top of representations produced by the frozen encoder $f_\theta$ and is trained independently from the adapters using cross-entropy loss on supervised domain annotations.
EMAR decouples routing and adapters training enabling independent extension and replacement of adapters without retraining the routing mechanism or previously learned domain-specific modules.
At inference time, as shown in Figure \ref{fig:distillationmodel}, EMAR infers the domain of the input query through the classifier and selects the corresponding domain-specific adapter, which is activated into the shared encoder. Then, the encoder processes queries and documents to produce the relevance score. In this way, EMAR requires a single forward pass regardless of the number of supported domains.

\subsection{Computational and Storage Efficency Properties}

From a system perspective, EMAR is designed to improve the efficiency of multi-domain retrieval.
In contrast to existing approaches that either require maintaining multiple full retrieval models or rely on a single shared model, EMAR evaluates the shared encoder exactly once per query, while domain specialization is introduced only through lightweight adapters. As shown in Table~\ref{tab:scaling}, this leads to a computation cost that is largely independent of the number of domains, while still preserving domain-specific behaviour through conditional computation.
Furthermore, EMAR supports incremental expansion: when a new domain $d_{N+1}$ is introduced, only a new adapter $A_{N+1}$ needs to be trained using in-domain supervision and distillation from a teacher model. The backbone encoder and existing adapters remain unchanged. Furthermore, since the routing function generalizes well across domains, it can optionally be extended when a new domain is added. 

\begin{table}[t]
\centering
\caption{Storage Costs across different strategies on $N$ domains. $P$ and $A$ are the number of parameters of the backbone and of a single adapter, respectively, and $A\ll P$.}
\label{tab:scaling}
\resizebox{\linewidth}{!}{%
\begin{tabular}{lcc}
\toprule
\textbf{Strategy} & \textbf{Param. Storage} & \textbf{New Domain Cost} \\
\midrule
Ensemble ($N$ models) & $N \cdot P$  & Full model retrain \\
Unified (1 model) & $P$ & Full model retrain \\
EMAR & $P + N \cdot A + A_{Routing}$ & One adapter and the classifier adapter (if needed) \\
\bottomrule
\end{tabular}%
}
\end{table}

\begin{table}[ht]
\centering
\caption{Experimental results on Academic Search and SE-PQA. $^*$ and $^\dagger$ denote significant improvements over S and ALL.}
\label{tab:results_academic_bi}
\resizebox{\linewidth}{!}{
\begin{tabular}{cl|c|c|c|c|c|c|c|c|c}
& & \multicolumn{4}{c|}{Academic Search} 
& \multicolumn{4}{c}{SE-PQA} \\ \cmidrule(lr){3-10}

 & \textbf{Model} & 
Comp Sci & Physics & 
Pol Sci & Psych & Apple & English & Gaming & SciFi \\

\midrule
\multirow{5}{*}{%
  \rotatebox[origin=c]{90}{%
    \parbox{1cm}{\centering
    Baseline}
  }
}
& BM25 & .225  & .269  & .241  & .239 & .276 & .506 & .474 & .521 \\
& DistilBERT (MS MARCO) & .231 & .273 & .254 & .257 & .291 & .535 & .502 & .530 \\
& BERT (MS MARCO) & .238 & .281 & .266 & .268 & .303 & .543 & .516 & .537 \\
& MPNet & .259 & .285 & .277 & .291 & .364 & .539 & . 570 & .629 \\
& MPNet-cQA & .244 & .289 & .267 & .276 & .361 & .590 & .571 & .629 \\


\midrule
\multirow{4}{*}{%
  \rotatebox[origin=c]{90}{%
    \parbox{1cm}{\centering
    MiniLM bi-encoder}
  }
}
& Fine-tuned (S)  & .\textbf{300} &   .\textbf{335} &  .295 & .\textbf{342} & \phantom{$^\dag$}.\textbf{413}$^\dag$ & \phantom{$^\dag$}.\textbf{644}$^\dag$ & \phantom{$^\dag$}.\textbf{627}$^\dag$  & \phantom{$^\dag$}.\textit{677}$^\dag$  \\
& Fine-tuned (ALL)  & .292  & .330  & .304  & .335 & .406  & .\textit{639}  & .615  & .664 \\
& EMAR LoRA  &   .\textit{293} & .\textit{332} & .\textit{305} & \phantom{$^\dag$}.\textit{341}$^\dag$ & .407  & .635  & .617  & \phantom{$^\dag$}.673$^\dag$ \\
& EMAR Houlsby & \phantom{$^\dag$}.\textbf{300}$^\dag$ & \phantom{$^\dag$}.\textbf{335}$^\dag$ & \phantom{$^\dag$}.\textbf{307}$^{*\dag}$ & \phantom{$^\dag$}.\textit{341}$^\dag$ & .\textit{409}  & \phantom{$^\dag$}.\textbf{644}$^\dag$  & \phantom{$^\dag$}.\textit{624}$^\dag$ & \phantom{$^\dag$}.\textbf{680}$^\dag$ \\

\midrule
\multirow{4}{*}{%
  \rotatebox[origin=c]{90}{%
    \parbox{1cm}{\centering
    MiniLM cross-encoder}
  }
}
& Fine-tuned (S)  & .\textbf{289} &   .\textbf{328} &  .\textbf{296} & .\textbf{339} &  \phantom{$^{\dag}$}.\textit{333}$^\dag$ & \phantom{$^{\dag}$}.\textit{579}$^\dag$ & \phantom{$^{\dag}$}.\textbf{554}$^\dag$ & \phantom{$^{\dag}$}.\textbf{586}$^\dag$ \\
& Fine-tuned (ALL)  & .273  & .307  & .292  & .312 & .323  & .567  & .542  & .\textit{570} \\
& EMAR LoRA  &   \phantom{$^\dag$}.276$^\dag$ & \phantom{$^\dag$}.311$^\dag$ & \phantom{$^\dag$}.\textit{293}$^\dag$ & \phantom{$^\dag$}.316$^\dag$ & \phantom{$^{\dag}$}.330$^{\dag}$ & \phantom{$^{\dag}$}.573$^{\dag}$ & .535   & .561  \\
& EMAR Houlsby & \phantom{$^\dag$}.\textit{283}$^\dag$ & \phantom{$^\dag$}.\textit{320}$^\dag$ & \phantom{$^\dag$}.\textbf{296}$^{*\dag}$ & \phantom{$^\dag$}.\textit{332}$^\dag$  & \phantom{$^{*\dag}$}.\textbf{340}$^{*\dag}$ & \phantom{$^{*\dag}$}.\textbf{585}$^{*\dag}$ & \phantom{$^{\dag}$}.\textit{553}$^\dag$ & \phantom{$^{\dag}$}.\textbf{586}$^\dag$  \\

\bottomrule
\end{tabular}
}
\end{table}

\section{Experimental Setup}
\label{sec:experimental_settings}

Our experiments aim to answer the following research questions: \textbf{RQ1}: Does EMAR's modular design achieve performance on par with domain-specific models? 
\textbf{RQ2}: Does EMAR provide a favorable effectiveness–efficiency trade-off?

\paragraph{Datasets.}
We train EMAR on two large-scale multi-domain benchmarks: (1)
Academic Search~\cite{bassani22amulti} is a large-scale citation prediction benchmark covering four domains (Computer Science, Physics, Political Science, Psychology), each containing approximately 5M documents. Queries are derived from paper titles, while cited papers are treated as relevant documents.
(2) SE-PQA~\cite{kasela24se-pqa} is a question-answering benchmark whose dataset is constructed from Stack Exchange. We rely on four communities (Apple, English, Gaming, SciFi), with 52k--100k training queries per domain. User questions are used as queries, and positively voted answers are treated as relevant documents. 
To evaluate EMAR on unseen domains, we rely on four BEIR datasets~\cite{thakur2021beir}, available through the ir-datasets package~\cite{MacAvaney_irdataset}. In detail, we use Natural Questions (NQ)~\cite{kwiatkowski2019natural} and HotpotQA (HQA)~\cite{yang2018hotpotqa} for open-domain Question Answering (QA) and Quora~\cite{thakur2021beir} and cqadupstack~\cite{hoogeveen2015cqadupstack} for duplicate-question retrieval tasks. Since cqadupstack is defined from Stack Exchange like SE-PQA, we select eight domains that are unseen during the training, i.e. Android, Gis, Mathematics, Programmers, Stats, Unix, Webmasters (Wm) and WordPress (Wp).  

\paragraph{Models and baselines.}
We rely on MiniLM~\cite{wang20minilm} as router and both student and teacher in bi-encoder and cross-encoder settings. Furthermore, for Academic Search, we train 
DistilBERT~\cite{sanh19distilbert} as router and student retriever with BERT-base~\cite{devlin-etal-2019-bert} as teacher.
We compare EMAR against different baselines: (i) BM25, a sparse lexical retrieval; (ii) RetroMAE~\cite{xiao2022retromae} and TCT-ColBERT~\cite{lin2020distilling}, which are bert-based dense retrievers trained through model distillation; (iii) DistilBERT- and BERT-MSMARCO, ANCE and TAS-B, which are dense retriever fine-tuned on MS MARCO~\cite{MSMARCO}; (iv) MPNet and MPNet-cQA~\cite{song2020mpnet}, which are dense retriever trained on a heterogeneous source of data, including citation prediction and question answering; (v) DESIRE-ME~\cite{kasela24desire} and SB-MOE~\cite{sokli2025mixture}, which are Mixture-of-Experts (MoE) approaches trained for question answering tasks;
(vi) MiniLM and DistilBERT fine-tuned independently for each training domain (S) or 
jointly fine-tuned across all training domains (ALL). 
Evaluating multiple model sizes and retrieval architectures allows us to assess EMAR across different effectiveness-efficiency trade-offs.
To assess the robustness of EMAR to the choice of adapter architecture, we evaluate two variants, i.e. Houlsby~\cite{houlsby19parameter} and LoRA~\cite{hu2022lora} adapters. 
These approaches introduce domain-specific parameters in different ways~\cite{lialin2023scaling,BRAGA_madakron}. Houlsby adapters add trainable bottleneck modules to the transformer architecture, increasing the number of parameters and the computational cost of inference. In contrast, LoRA learns low-rank updates to the model weights, which can be merged into the backbone at inference time, avoiding additional inference computations. We select these two approaches to investigate the trade-off between retrieval effectiveness and computational efficiency in EMAR. A broader comparison with other parameter-efficient approaches, including prefix tuning~\cite{zhang2023towards}, is left for future work.
 
\begin{table}[ht]
\centering
\scriptsize
\caption{Experimental results on Academic Search with DistilBERT. Notation as per Table \ref{tab:results_academic_bi}.}
\label{tab:results_academic_d}
\resizebox{0.7\linewidth}{!}{
\begin{tabular}{c|c|c|c|c|c}

 \textbf{Model} & 
Comp Sci & Physics & 
Pol Sci & Psych \\

\midrule
 BM25 & .225  & .269  & .241  & .239  \\
DistilBERT (MS MARCO) & .231 & .273 & .254 & .257  \\
BERT (MS MARCO) & .238 & .281 & .266 & .268  \\
 BERT (teacher) &  .314  & .364  & .303  & .358 \\

\midrule
 Fine-tuned (S)  & .304  & \phantom{$^\dag$}.\textbf{356}$^\dag$ &   .300  & \phantom{$^\dag$}.344$^\dag$ \\
 Fine-tuned (ALL)  & .300  & .345  & .306 & .337 \\
 EMAR LoRA & \phantom{$^{*\dag}$}.\textit{307}$^{*\dag}$ & .345 & \phantom{$*\dag$}.\textit{312}$^{*\dag}$ & \phantom{$^\dag$}.\textit{345}$^\dag$ \\ 
 EMAR Houlsby & \phantom{$^{*\dag}$}.\textbf{314}$^{*\dag}$  & \phantom{$^\dag$}.\textit{352}$^\dag$ & \phantom{$^{*\dag}$}.\textbf{319}$^{*\dag}$ & \phantom{$^{*\dag}$}.\textbf{349}$^{*\dag}$ \\
\bottomrule
\end{tabular}
}
\end{table}

\paragraph{Training and Evaluation Details}

We train all full retrieval models on a NVIDIA A100 for 40 epochs with a batch size of 128, learning rate $2\times10^{-5}$ and margin 0.1. Adapter-based variants are trained in the same setting but with a learning rate of $5\times10^{-7}$, while the routing classifier is trained independently for 10 epochs with learning rate $10^{-4}$.  
During inference, following prior works~\cite{bassani22amulti,kasela24se-pqa,thakur2021beir}, we adopt a standard re-ranking pipeline: BM25 first retrieves the top-1000 candidate documents for Academic Search and the top-100 candidates for SE-PQA and BEIR, which are subsequently re-ranked by the neural retrievers. For all datasets, we use nDCG@10~\cite{dunn1961multiple} to evaluate the performance of the proposed approach.
For significance testing, we use the paired t-test with $p<0.01$ and Bonferroni correction~\cite{dunn1961multiple}. Finally, we make our code publicly available.\footnote{\url{https://github.com/DetectiveMB/EMAR}}   



\section{Results}
\label{sec:results}

In this Section, we show the results in nDCG@10 of our evaluation in Tables~\ref{tab:results_academic_bi},~\ref{tab:results_academic_d},~\ref{tab:ablation_routing} and~\ref{tab:results_generalization}, where, for each retrieval architecture, we report the best results in \textbf{bold} and the second-best in \textit{italic}.  

\paragraph{RQ1: Does EMAR’s modular design achieve performance on par with domain-specific models?}

We first investigate whether EMAR can match domain-specific retrievers across heterogeneous domains. The results in Tables~\ref{tab:results_academic_bi} show that domain-specific models (S) generally outperform unified training (ALL), with performance gaps reaching up to 3.2\% in nDCG@10 across both Academic Search and SE-PQA. The largest differences are observed in topically coherent domains such as Physics, Gaming, and SciFi, while smaller gaps appear in more heterogeneous domains such as Political Science and English.
In contrast, EMAR consistently closes this gap. Across both bi-encoder and cross-encoder settings, 
EMAR Houlsby achieves performance on par or even better than domain-specific models, with significant improvements in five of eight domains, and significantly outperforms unified models by up to 4.7\% nDCG@10.
EMAR LoRA shows slightly lower but still competitive performance. 
Since EMAR Houlsby with MiniLM as bi-encoder achieves the best overall performance, we henceforth refer to it simply as EMAR, unless otherwise specified.
\begin{table}[t]
\centering
\caption{Impact in nDCG@10 of the routing mechanism. * denotes statistical significant differences over random.}
\resizebox{\linewidth}{!}{
\begin{tabular}{l|cccc|ccccc}

& \multicolumn{4}{c|}{Academic Search} 
& \multicolumn{5}{c}{SE-PQA} \\ \cmidrule(lr){2-10}
Router & 
Comp Sci & Physics & 
Pol Sci & Psych & Apple & English & Gaming & SciFi 
& OOD \\
\midrule
Classifier & \phantom{*}.300* & .335 & \phantom{*}.307* & .341 & \phantom{*}.409* & \phantom{*}.644*  & \phantom{*}.624* & \phantom{*}.680* & 578  \\
Random & .290 & .330 & .304 & .335 & .323 & .567 & .542 & .570  & 569 \\
Oracle  & .300 & .335 & .307 & .341 & .409 & .644  & .624 & .680 & - \\
\end{tabular}
}
\label{tab:ablation_routing}
\end{table}
To study the impact of the classifier, we analyse the performance of the routing function, which achieves high accuracy across most domains, reaching 89\% on Academic Search and 99\% on SE-PQA, with the main exception being Political Science (63\%). This lower performance can be largely attributed to strong topical overlap with other academic domains. 
To better understand, we manually inspect several misclassified queries in the Political Science domain, such as ``Grazing behaviour and associations with obesity, eating disorders, and health-related quality of life in the Australian population,'' which is from Political Science but is classified by the router as Psychology. We leave as future works a detailed analysis of routing classifier in out-of-domain datasets. Importantly, as shown in Table \ref{tab:ablation_routing}, such ambiguities do not negatively impact overall retrieval performance: the performance of EMAR with the learned router are on par with EMAR when we select the correct domain a priori (i.e. an Oracle). Furthermore, we show that the learned router achieves superior performance over random selection in both training and unseen (OOD) domains, underlying the need of the learned domain classifier.
Table~\ref{tab:results_generalization} further shows that EMAR and the frozen learned router generalise effectively on unseen domains from BEIR.
In open-domain QA, EMAR consistently outperforms MiniLM ALL and achieves performance on par with MPNet, showing that the learned adapters can be effectively reused for semantically related but unseen distributions. Similar trends are observed on duplicate-question tasks, where EMAR achieves the best or second-best performance in seven over eight domains. Finally, in Table~\ref{tab:results_moe}, we compare EMAR with two MoE-based retrieval approaches, DESIRE-ME and SB-MOE, on HotpotQA and Natural Questions (NQ). EMAR achieves competitive results despite being evaluated in a zero-shot setting, whereas the MoE-based approaches are trained on these datasets. In particular, EMAR outperforms several fine-tuned baselines and achieves better performance than COCO-DR on HotpotQA, showing EMAR as a promising approach for multi-domain adaptation. On NQ, however, COCO-DR and Contriever outperform EMAR, with statistically significant differences. 
To concretely answer RQ1, these results show that EMAR effectively learns domain-specific knowledge without requiring model duplication. 

\begin{table}[t]
\centering
\caption{Zero-Shot results on BEIR. Symbol * denotes statistically significant difference between the best approach and the second best.}
\resizebox{\linewidth}{!}{%
\begin{tabular}{c|c|c|c|c|c|c|c|c|c|c|c}
Model &
HQA & NQ  &
Quora & Andr. & Gis & Math. & Prog. & Stats & Unix & Wm & Wp  \\
\midrule

BM25  & .586 & .281  & .770 & .338 & .258 & .194 & .250 & .228 & .241 & .265 & .192 \\ 
 ANCE & .451 & .443 & .852 & .387 & .259 & .179 & .316 & .262 & .268 & .297 & .237 \\
 TAS-B & .584 & .463 & .835 & .404 & .296 & .209 & .312 & .281 & .291 & .311 & .241 \\
 TCT-ColBERT & .630 & .331 & .777 & .342 & .267 & .207 & .275 & .240 & .253 & .269 & .209 \\
 RetroMAE & .603 & .362 & .842 & 371 & .278 & .214 & .280 & .245 & .276 & .299 & .217 \\ 
 DistilBERT & .637 & .359 & .836 & .365 & .271 & .209 & .285 & .245 & .266 & .293 & .216 \\
 BERT & .\textbf{643} & .367 & .841 & .371 & .281 & .213 & .290 & .256 & .273 & .296 & .224 \\
 COCO-DR & .604 & .\textbf{502} & .\textbf{867} & .\textbf{438} & \phantom{*}.\textbf{342}* & .233 & .\textit{332} & .\phantom{*}\textbf{318}* & .\textbf{327} & .\textbf{341} & .\textbf{275} \\
 Contriever & .638 & .\textit{497} & .\textit{865} & .\textit{426} & .302 & .235 & \phantom{*}.\textbf{357}* & .\textit{301} & .\textit{326} & .\textit{339} & .253 \\
 MPNet dense & .635 & .378 & .853 & .\textit{418} & .323 & .\textbf{255} & .322 & .282 & .314 & .331 & .249 \\
 MPNet-cQA & .\textit{641} & .383 & .851 & .407 & .318 & .\textit{250} & .316 & .287 & .309 & .328 & .251 \\
\midrule

Minilm (ALL)  & .630  &  .364 & .848 & .416 & .324 &  .239 & .320 & .272 & .315 & .322 & .\textit{264}  \\
 Minilm EMAR Houlsby & .633 & .376    & .854 & .425 & .\textit{330} &  .242 & .320 & .282 & .319 & .329 & .258 \\
\end{tabular}%
}
\label{tab:results_generalization}
\end{table}

\paragraph{RQ2: Does EMAR provide a favorable effectiveness–efficiency trade-off?}
EMAR applies knowledge distillation from domain-specific teachers, so its training cost is on par with the single-domain models. However, EMAR is more efficient in the number of parameters saved while introducing more inference costs.
Table~\ref{tab:per_model_energy} shows the storage cost in terms of model parameters and the inference cost per query, measured in GFLOPs~\cite{calflops} and CO$_2$ emissions~\cite{benoit_courty_2024_11171501}. Domain-specific models scale poorly across domains because each domain requires a separate copy of the retrieval backbone. For example, the DistilBERT ensemble (S) requires 264M parameters for four domains, corresponding to four independently stored 66M-parameter backbones. In contrast, EMAR shares a single frozen backbone across domains and the classifier and stores only lightweight domain-specific adapters. With LoRA, EMAR requires 132.7M parameters in total, representing approximately a 75\% reduction in storage compared to the domain-specific ensemble.
Furthermore, EMAR does not increase the per-query inference cost of the retrieval models when using LoRA. The increased overall computational costs are due to the application of the query classifier. 
In contrast, Houlsby adapters require additional computations, resulting in a higher inference cost while achieving superior retrieval performance. To finally answer RQ2, EMAR with LoRA provides a favourable effectiveness-efficiency trade-off by reducing storage requirements while achieving performance on par or even better to those of domain-specific models.

\begin{table*}[t]
\centering
\caption{Experimental results on NQ and HotpotQA compared with recent MoE-based approaches. Symbol * indicates a statistically significant difference over fine-tuned (FT).}
\label{tab:results_moe}
\resizebox{\linewidth}{!}{
\begin{tabular}{l|c|ccc|cccc|cc|cc|cc|c}
\toprule
 & BM25 & \multicolumn{3}{c}{COCO-DR} & \multicolumn{4}{c}{Contriever} & \multicolumn{2}{c}{Tiny-BERT} & \multicolumn{2}{c}{BERT-small} & \multicolumn{2}{c}{BERT-base} & MiniLM \\
\midrule
Training
 & / & Base & FT & DESIRE-ME
 & Base & FT & DESIRE-ME & SB-MOE
 & FT & SB-MOE
 & FT & SB-MOE
 & FT & SB-MOE
 & EMAR Houlsby \\
\midrule
HotpotQA
 & .568 & .604 & .608 & \phantom{*}.614*
 & .638 & .657 & \phantom{*}.648* & \phantom{*}.667*
 & .243 & \phantom{*}.252*
 & .310 & \phantom{*}.317*
 & .444 & .431
 & .633 \\
NQ
 & .281 & .502 & .501 & .526
 & .497 & .497 & .559 & .416
 & .261 & \phantom{*}.271*
 & .215 & \phantom{*}.218*
 & .350 & \phantom{*}.356*
 & .376 \\
\bottomrule
\end{tabular}
}
\end{table*}

\begin{table}[t]
\centering
\caption{Number of parameters for four domains and Query-wise inference FLOPs and CO$_2$ of EMAR. 
}
\label{tab:per_model_energy}
\resizebox{\linewidth}{!}{%
\begin{tabular}{lcccc}
\toprule
Model & Memory Params. (M) & Activeted Params./query (M) & GFLOP/query & CO$_2$ (10$^{-5}$ kg)  \\
\midrule
DistilBERT$_{\text{S}}$         & 264.0 & 66.0 & 11.18 &  5.1  \\
DistilBERT$_{\text{ALL}}$       & 66.0 & 66.0 & 11.18 &  1.3 \\
EMAR-DistilBERT$_{\text{LoRA}}$     & 66.8 & 132.0 & 22.4 & 2.7      \\
EMAR-DistilBERT$_{\text{Houlsby}}$     & 212.6 & 293.2 & 28.2 & 2.9 \\
\midrule
MiniLM$_{\text{S}}$             & 90.8 & 22.7 & 2.87 &    5.2  \\
EMAR-MiniLM$_{\text{LoRA}}$     & 23.1 & 45.4  & 5.74    & 2.6 \\
EMAR-MiniLM$_{\text{Houlsby}}$         & 78.4 & 58.6 & 7.74 & 2.8  \\
\bottomrule
\end{tabular}%
}
\end{table}

\section{Conclusions}
\label{sec:conclusions}

We propose EMAR, a modular IR system for efficient multi-domain neural retrieval.
Experiments across seventeen domains and three retrieval architectures show that EMAR achieves retrieval effectiveness comparable to domain-specific models while reducing computational overhead and storage.

\section{Acknowledgments}
Braga acknowledges the CINECA award under the ISCRA initiative, for the availability of high-performance computing resources and support.

\section*{Declaration on Generative AI}

 During the preparation of this work, the author(s) used GPT-4 in order to: Grammar and spelling check. After using these tool(s)/service(s), the author(s) reviewed and edited the content as needed and take(s) full responsibility for the publication’s content.

\bibliography{sample-ceur}


\end{document}